\documentclass{pasj00}

\begin{document}
\SetRunningHead{S.\ Yamauchi et al.}{X-Ray Emission from 
Supernova Remnants Observed in the ASCA Galactic Plane Survey}
\Received{2007/10/22}
\Accepted{yyyy/mm/dd}

\title{X-Ray Emission from Supernova Remnants Observed in the ASCA Galactic Plane Survey}

\author{Shigeo \textsc{Yamauchi}}%
\affil{Faculty of Humanities and Social Sciences, Iwate University, 
3-18-34 Ueda, Morioka, Iwate 020-8550}
\email{yamauchi@iwate-u.ac.jp}

\author{Masaru \textsc{Ueno}} 
\affil{Department of Physics, Faculty of Science, Tokyo Institute
 of Technology, 2-12-1 Oo-okayama, Meguro-ku, Tokyo 152-8551}

\author{Katsuji \textsc{Koyama}}
\affil{Department of Physics, Graduate School of Science, Kyoto University, 
Sakyo-ku, Kyoto 606-8502}

\and
\author{Aya \textsc{Bamba}}
\affil{Institute of Space and Astronautical Science (ISAS),
Japan Aerospace Exploration Agency (JAXA), 
3-1-1 Yoshinodai, Sagamihara, Kanagawa 229-8510}

%


\maketitle

\begin{abstract}
X-ray images and spectra of 5 cataloged supernova remnants (SNRs), 
G12.0$-$0.1, G346.6$-$0.2, G348.5$+$0.1, G348.7$+$0.3, and G355.6$-$0.0,
observed in the ASCA galactic plane survey are presented.
The sizes of X-ray emission from G12.0$-$0.1, G348.5$+$0.1, G348.7$+$0.3,
and G355.6$-$0.0 are 
comparable to their radio structures,
while that of G346.6$-$0.2 is 
smaller than the radio structure.
The X-ray spectra of all of the SNRs 
were heavily absorbed by interstellar matter 
with $N_{\rm H}>10^{22}$ cm$^{-2}$.
The spectrum of G355.6$-$0.0 
exhibited emission lines, 
indicating that the X-ray emission has a thin thermal plasma origin, 
and was well represented by two-temperature thin thermal emission model.
On the other hand, 
no clear emission line features were found in the spectra of the others
and the spectra could be represented by either 
a thin thermal emission model or a power-law model.
\end{abstract}

\section{Introduction}

Supernovae (SNe) and supernova remnants (SNRs) 
are very important objects, because they are main sites 
of heavy element production and of acceleration of high-energy
particles in galaxies.
X-ray observations of SNRs are useful for performing
plasma diagnostics and for searching for synchrotron X-rays from 
high-energy electrons.

Many survey observations of galactic SNRs have been 
carried out mainly in the
radio band and 265 SNRs have been cataloged so far \citep{Green2006}. 
The Einstein Observatory detected 
X-rays from about 40 SNRs (e.g., \cite{Seward1990}),
while the ROSAT all-sky survey detected X-rays from many SNRs 
including new SNRs (e.g., \cite{Pfeffermann1991,Pfeffermann1996,Voges1999}).
However, X-rays in the low energy band are
absorbed by the interstellar medium, and hence soft X-ray observations 
can detect X-rays from only nearby SNRs on the galactic plane.
On the other hand, although hard X-rays can observe SNRs even in the 
galactic plane, hard X-ray detectors without imaging capability could 
not find faint X-ray emission from SNRs on the galactic plane
where there are many bright X-ray sources containing a compact object.
The imaging capability in the hard X-ray band is 
essential to search for X-ray emission from SNRs located in the galactic disk.
 
%
\begin{table*}[t]
\caption{The properties of SNRs. }
\begin{center}
\begin{tabular}{lccccccc} \hline \\ [-6pt]
Name &  Position$^a$ & Size$^a$ & Type$^{a, b}$ & $F_{\rm 1GHz}^{a, c}$ & 
$\Gamma_{\rm Radio}^{a, d}$  & $F_{\rm TeV}^e$ & $\Gamma_{\rm TeV}^f$\\
     & (RA, Dec)$_{\rm J2000}$ & (arcmin) &      & (Jy) &  & (erg s$^{-1}$ cm$^{-2}$) &\\
\hline \\[-6pt]
G12.0$-$0.1  & (18$^{h}$12$^{m}$11$^{s}$, $-$18$^{\circ}$37$^{'}$) & 7? & ? & 3.5 & 0.7 & --- & --- \\   
G346.6$-$0.2 & (17$^{h}$10$^{m}$19$^{s}$, $-$40$^{\circ}$11$^{'}$) & 8 & S & 8? & 0.5? & --- & --- \\
G348.5$+$0.1 & (17$^{h}$14$^{m}$06$^{s}$, $-$38$^{\circ}$32$^{'}$) & 15 & S & 72& 0.3 & 2.3$\times10^{-12}$ & 2.30 \\
G348.7$+$0.3 & (17$^{h}$13$^{m}$55$^{s}$, $-$38$^{\circ}$11$^{'}$) & 17? & S & 26 & 0.3 & 1.2$\times10^{-12}$ & 2.65\\
G355.6$-$0.0 & (17$^{h}$35$^{m}$16$^{s}$, $-$32$^{\circ}$38$^{'}$) & 8$\times$6 & S & 3? & ? & --- & --- \\
\hline\\
\end{tabular}
\end{center}
$^a$ \citet{Green2006} and references therein.\\
$^b$ S: shell-type and C: composite-type.\\
$^c$ Radio flux at 1 GHz.\\
$^d$ Spectral index in the radio band.\\
$^e$ Unabsorbed energy flux in the 1--10 TeV energy band 
(Aharonian et al. 2008a, 2008b).\\
$^f$ Power-law index of the TeV $\gamma$-ray spectrum 
(Aharonian et al. 2008a, 2008b).\\
\end{table*}
%
\begin{table*}[t]
\caption{The observation logs. }
\begin{center}
\begin{tabular}{lcccc} \hline \\ [-6pt]
Name &  Observation date & Pointing position & 
$\Delta\theta^{a}$ & Exposure time \\
     &                   & (RA, Dec)$_{\rm J2000}$ & (arcmin) & (ks) \\
\hline \\[-6pt]
G12.0$-$0.1  & 1996 April 9     & 
(18$^{h}$11$^{m}$53.9$^{s}$, $-$18$^{\circ}$32$^{'}$03.1$^{''}$) & 6.4 & 6.3\\ 
G346.6$-$0.2 & 1996 September 3 & 
(17$^{h}$09$^{m}$34.1$^{s}$, $-$39$^{\circ}$59$^{'}$35.9$^{''}$) & 14.3 & 8.2\\
G348.5$+$0.1 & 1996 September 12& 
(17$^{h}$15$^{m}$38.1$^{s}$, $-$38$^{\circ}$22$^{'}$37.2$^{''}$) & 20.3 & 7.8\\
G348.7$+$0.3 & 1996 September 12& 
(17$^{h}$15$^{m}$38.1$^{s}$, $-$38$^{\circ}$22$^{'}$37.2$^{''}$) & 23.3 & 7.8\\
G355.6$-$0.0 & 1996 March 16    & 
(17$^{h}$34$^{m}$57.5$^{s}$, $-$32$^{\circ}$34$^{'}$54.5$^{''}$) & 5.0 & 9.8\\
\hline\\
\end{tabular}
\end{center}
$^a$ Separation angle between the SNR position (see table 1) and the FOV center.\\
\end{table*}

ASCA was the first satellite carrying X-ray detectors with 
imaging capability in the 0.4--10.0 keV energy band \citep{Tanaka1994}.
We performed survey observations on the galactic plane with ASCA.
The ASCA
galactic plane survey (AGPS) covered all the Galactic 
inner disk ($| l | < 45^{\circ}$ and $| b | < 0.4^{\circ}$) and the 
Galactic center region ($| l |<2^{\circ}$ and $| b |<2^{\circ}$)
with successive pointing 
observations of about 10 ks exposure \citep{Yamauchi2002}.
One of the major objectives of AGPS is to search for
X-ray emission from cataloged radio SNRs and to investigate their properties.
AGPS detected X-ray emissions from $\sim$30
cataloged SNRs, including 15 SNRs undetected in the X-ray band before AGPS. 
\citep{Yamauchi1998, Sugizaki2001, Sakano2002, Yamauchi2002}.
Results of several SNRs observed in AGPS have been 
reported by several authors
(G352.7$-$0.1: \cite{Kinugasa1998}; 
G359.0$-$0.9: \cite{Bamba2000}; 
G359.1$-$0.5: \cite{Yokogawa2000}; 
G359.1$+$0.9 and G1.9$+$0.3: \cite{Sakano2002}; 
G344.7$-$0.1: \cite{Yamauchi2005}).
Furthermore, 
G15.9$+$0.2, G16.7$+$0.1, G338.3$-$0.0, and G349.7$+$0.2 have been
observed with ASCA, Chandra, and XMM-Newton 
after the X-ray detection in AGPS, 
and the results have been reported 
(G15.9$+$0.2: \cite{Reynolds2006}; G16.7$+$0.1: \cite{Helfand2003}; 
G338.3$-$0.0: \cite{Funk2007}; G349.7$+$0.2: 
\cite{Slane2002,Lazendic2005}). 
In this paper, we report on results of other 5 SNRs, 
G12.0$-$0.1, G346.6$-$0.2, 
G348.5$+$0.1, G348.7$+$0.3, and G355.6$-$0.0,
obtained from the AGPS data.
The properties of the SNRs are listed in table 1.

\section{Observations and Data Reduction}

AGPS was performed in 1996--1999. 
ASCA carried two Solid-state Imaging 
Spectrometers (SIS0, SIS1) and 
two Gas Imaging Spectrometers (GIS2, GIS3) placed at the focal planes 
of four thin foil X-ray Telescopes (XRT).
Each SIS camera,  consisting of 
4-CCD chips, was operated in 4-CCD mode or 2-CCD mode.
In order to avoid telemetry overflow, 
the lower discriminator was set to be 0.7 keV, and hence
the effective energy range of the SIS was 0.7--10 keV.
On the other hand, 
the GIS was operated in PH mode with the 0.7--10 keV band. 
The two sets of the SIS covered 22$'\times$22$'$, 
while the field of view (FOV) of the GIS was $50'$
in diameter.
Details of ASCA and the instruments are given in separate papers 
(ASCA satellite: \cite{Tanaka1994}; XRT: \cite{Serlemitsos1995}; 
SIS: \cite{Burke1991}; GIS: \cite{Makishima1996, Ohashi1996}).


The SIS data quality exhibited significant degradation due mainly to
the particle irradiation 
\citep{Dotani1995, Yamashita1997}.
The most serious problem for the present SNR observation is the increase
of pixel-to-pixel fluctuation of the dark current 
(Residual Dark Distribution, RDD), which makes the energy
resolution worse.
Therefore, we concentrate on results using the GIS data.

We excluded the data obtained at the South
Atlantic Anomaly, during the earth occultation, at the low elevation
angle from the earth rim of $<$ 5$^{\circ}$, 
and in the high background regions at low geomagnetic cut off
rigidities of $<$ 8 GV.  
We also applied a rise-time discrimination technique 
to reject particle events.
The observation logs are listed in table 2.

\section{Analysis and Results}

%
\begin{table*}[t]
\caption{The best-fit parameters of spectral analysis$^a$.}
\begin{center}
\begin{tabular}{lcccccc} \hline \\ [-6pt]
Name & Model$^b$ &
 $kT$ /$\alpha^c$ & $N_{\rm H}^d$  & $\chi^2$ (d.o.f.) & $F_{\rm obs.}^e$& $F_{\rm unabs.}^f$ \\
\hline \\[-6pt]
G12.0$-$0.1  &1T & 1.6 ($>$0.7) & 6.5 (1.0--13.8) & 5.24 (12) &0.79 &6.4\\   
            & PL & 3.1 (1.2--6.3) & 5.9 (1.4--14.7) & 4.31 (12)&0.91&8.7 \\  
G346.6$-$0.2 &1T & 1.6 (0.4--4.7) & 2.0 (0.4--5.5) & 11.00 (18) &1.3&5.6\\   
             &PL & 3.7 ($>$1.7)   & 2.6 (0.4--10.8) & 12.06 (18) &1.3&18\\   
G348.5$+$0.1 &1T & 2.2 (1.8--3.2) & 2.0 (1.4--2.5) &34.98 (40)& 6.6&18\\    
             &1Tv$^g$ & 2.8 (2.1--4.4) & 1.7 (1.2--2.3) &26.68 (39)& 6.9&15\\
             &PL & 3.0 (2.4--3.6) & 2.4 (1.8--3.2) &32.84 (40)& 6.7&36\\     
G348.7$+$0.3$^h$ &1T & 1.6 (0.9--2.6)& 2.7 (1.4--5.1)& 18.49 (17)& 2.5 & 12\\
 &  PL & 4.1 (2.8--6.1)& 4.0 (2.0--7.3) & 17.76 (17)& 2.5 & 82\\  
G355.6$-$0.0 & 1T & 1.0 (0.6--1.5)& 4.0 (2.8--5.4) & 27.55 (27)&0.81& 17\\
             & 2T & 0.7 (0.3--1.0)& 4.4 (3.2--6.6) & 21.03 (25)&0.58& 40\\
             &    & 7.5 ($>$ 1.7) & 4.4$^i$        & --- & 0.48& 0.94\\    
             & PL & 4.5 (3.2--6.2) & 4.4 (2.7--6.7)& 32.95 (27) & 0.90& 50\\    
\hline\\
\end{tabular}
\end{center}
$^a$ Values in parentheses show 90\% confidence intervals.\\
$^b$ PL: power-law model, 
1T: single-temperature MEKAL model with the Solar abundances, 
1Tv: single-temperature MEKAL model with variable abundances, 
2T: 2-temperature MEKAL model with the Solar abundances.\\
$^c$ Temperature (keV)/photon index.\\
$^d$ Hydrogen column density in unit of 10$^{22}$ cm$^{-2}$.\\
$^e$ Observed X-ray flux in the 0.5--10 keV energy band calculated using 
the best-fit parameters. 
The unit is 10$^{-12}$ erg s$^{-1}$ cm$^{-2}$.\\
$^f$ $N_{\rm H}$-corrected X-ray flux in the 0.5--10 keV energy band 
calculated using the best-fit parameters. 
The unit is 10$^{-12}$ erg s$^{-1}$ cm$^{-2}$.\\
$^g$ The Si abundance was set to be free; Si=4.3$^{+6.1}_{-2.3}$ Solar.\\
$^h$ ASCA observed only the southern 
part of the SNR because the SNR was located 
near to the edge of the GIS FOV.\\
$^i$ The same $N_{\rm H}$ value as that for the low-temperature component
was assumed.\\
\end{table*}

The survey area contained $\sim$60 
SNRs cataloged before AGPS \citep{Green2006} 
and AGPS detected X-ray emission at the positions of 
$\sim$30 SNRs, including 15 SNRs undetected in the X-ray band before AGPS.
Here, we report on the X-ray properties of 5 SNRs 
of which X-ray properties have not been reported.
The X-ray images in the 0.7--10.0 keV energy band  for GIS2 and GIS 3 were 
made and 
the data obtained with GIS2 and GIS3 were combined.
The X-ray spectrum was extracted from the region including the SNR.
The background spectrum was extracted from the
source free region in the same FOV.
In order to maximize photon statistics, the data obtained with GIS2 and GIS3 
were combined.

Several studies reported that there were some differences in the energy scales
between the SIS and the GIS and that a gain adjustment of 2--3 \% was required 
for the GIS analysis \citep{Miyata1996, Yamauchi1999, Slane2002, Yamauchi2005}.
However, since the photon statistics were limited due to short exposure 
time and low source flux, 
the precise gain adjustment was impossible for the present analysis.
Taking account of these facts, 
we used two simple models for a spectral analysis;
a thin thermal emission model from a plasma in the 
collisional ionization equilibrium 
(MEKAL model; \cite{Mewe1985,Liedahl1995,Mewe1995}) and a power-law model,
and did not apply a more detailed model such as a non-ionization equilibrium
model. 
The abundance table were taken from \citet{Anders1989}, while
the cross sections of photoelectric absorption were taken from 
\citet{Morrison1983}.
We, at first, assumed the abundances to be solar, but 
in some cases we set the metal abundances to be free.
Results of the spectral analysis are summarized in table 3.
X-ray properties for each SNR are described in the following section.

\subsection{G12.0$-$0.1}

\begin{figure*}
  \begin{center}
    \FigureFile(70mm,70mm){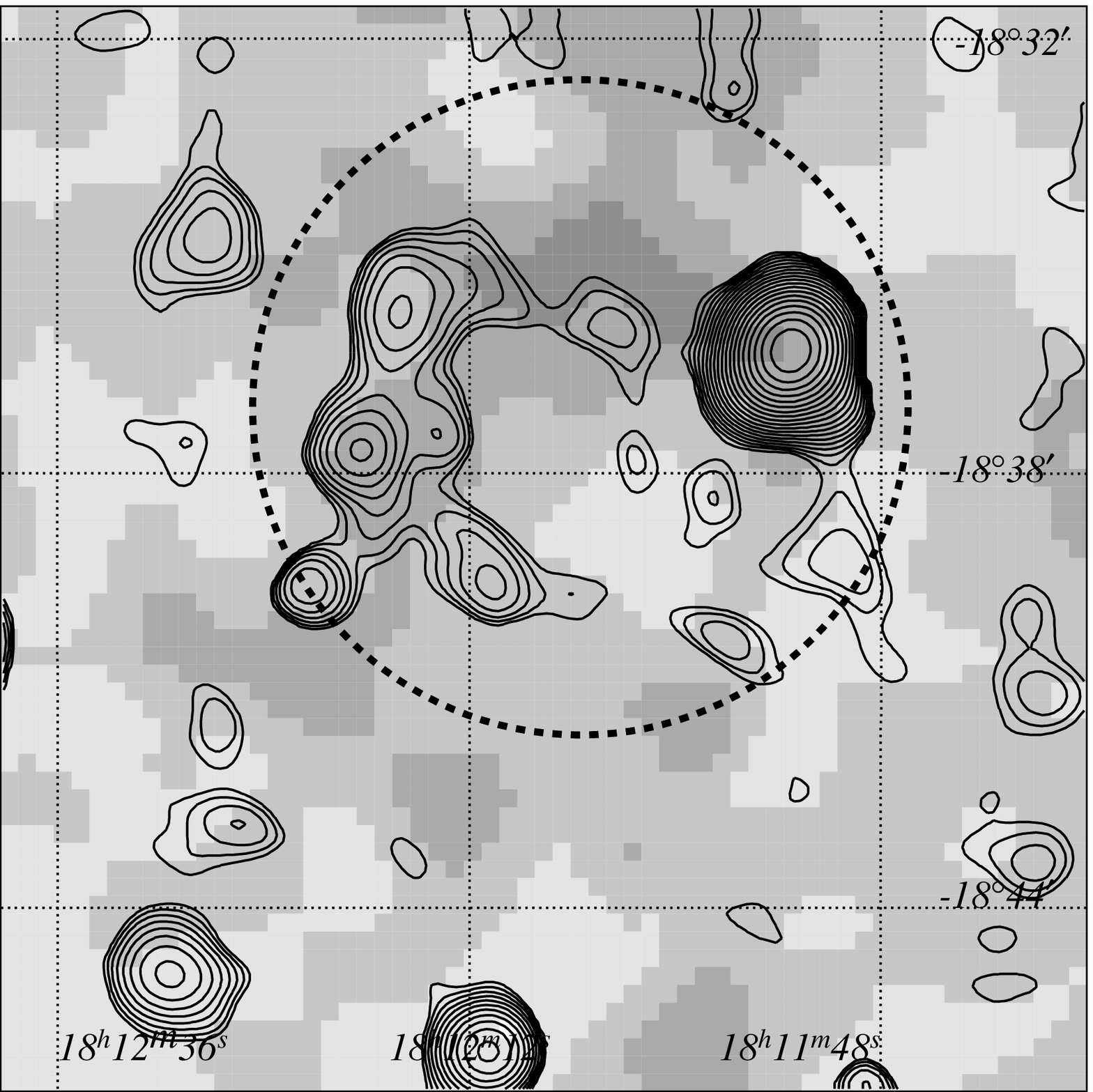}
    \FigureFile(90mm,90mm){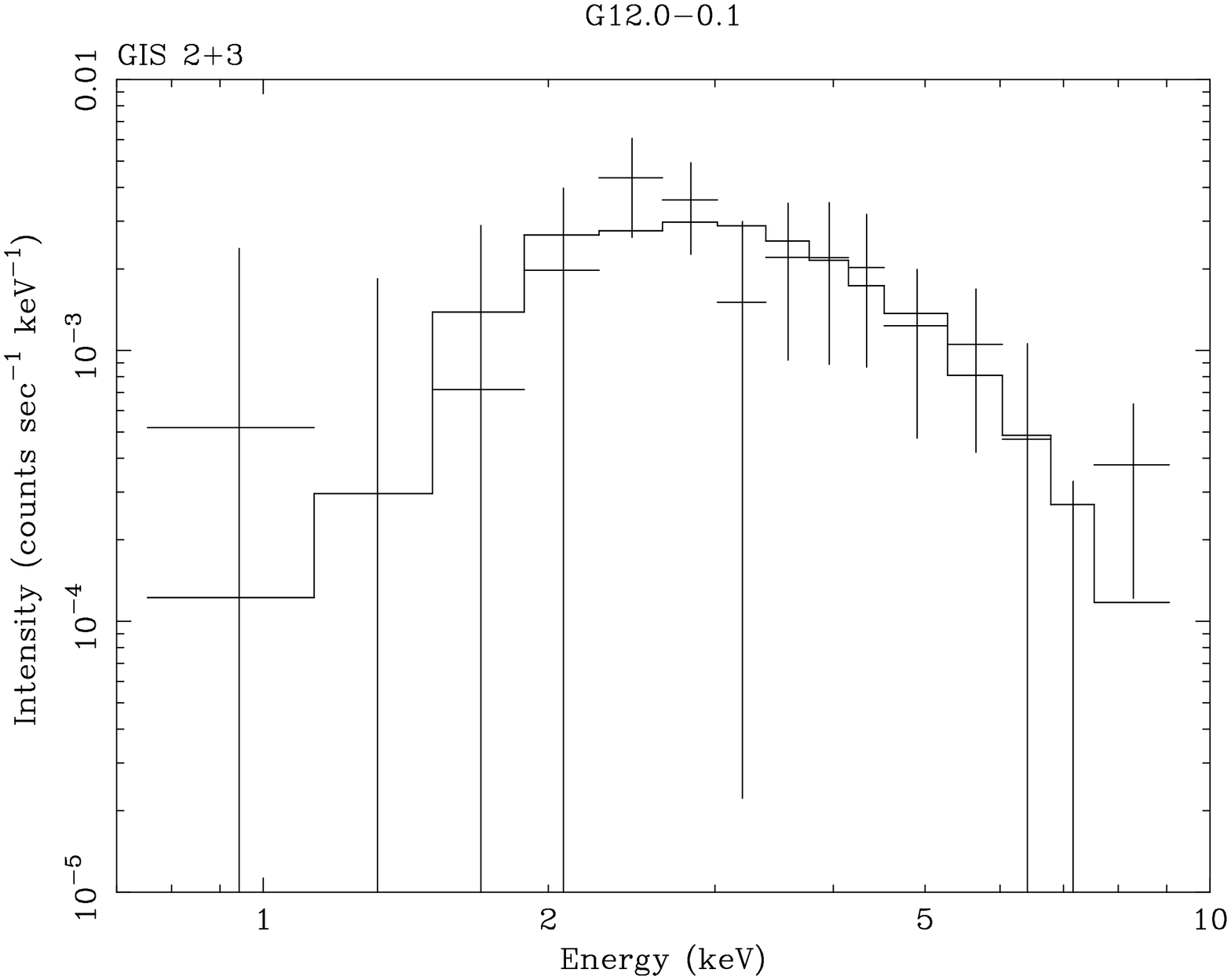}
  \end{center}
  \caption{
Left: GIS 2+3 image of G12.0$-$0.1 obtained in the 0.7--10.0 keV energy band 
(gray scale).
The X-ray image was smoothed with a Gaussian distribution of $\sigma$ = 30$''$.
The coordinates are J2000.
The intensity level are linearly spaced by 0.2 counts pixel$^{-1}$.
The contour shows a radio intensity map at 1.4 GHz using the NRAO 
VLA Sky Survey (NVSS) \citep{Condon1998}.
The dotted line shows the region from which the X-ray spectrum was extracted.
Right: GIS spectrum of G12.0$-$0.1 (the crosses) 
and the best-fit power-law model (the histogram).
}\label{fig1}
\end{figure*}

X-ray detection of G12.0$-$0.1 
in AGPS 
and results of a power-law model fit were
reported by \citet{Sugizaki2001}.
The radio structure shows an incomplete shell, 
where only the eastern half is visible (figure 1, e.g., \cite{Condon1998}).
The bright source at the western part in the radio band image is a thermal 
radio source 
containing an ultracompact HII region 
\citep{Downes1980,Becker1994}.
The X-ray image (the gray scale in figure 1) 
displayed weak X-ray emission.
However, no clear X-ray peak was seen at the position of 
the western bright radio source.

The X-ray spectrum was extracted from a 4.5$'$ radius circle including 
the arc-like structure in the radio band.
The background spectrum was extracted from an annular region between 
9$'$ and 4.5$'$ radius, but
X-ray photons from the 3$'$ radius circle around AX J181213$-$1842 
\citep{Sugizaki2001} and two local peaks at the north 
of G12.0$-$0.1, not listed in \citet{Sugizaki2001}, 
were excluded.
The X-ray counts from the source region in the 0.7--10 keV energy band 
were 568 (GIS2$+$3), 
while the background counts in the same region size and in the same energy 
band were estimated to be 446 (GIS2$+$3).
Since the photon statistics were limited,
the X-ray spectrum was well explained by either a MEKAL
model or a power-law model.
The X-ray spectrum and the best-fit power-law model are shown in figure 1.
The spectral parameters obtained from the power-law model 
fit were consistent with those in \citet{Sugizaki2001} within errors.
The differences in the best-fit spectral parameters 
between \citet{Sugizaki2001} and our analysis 
would be due to the differences in the source region: 
in \citet{Sugizaki2001}, the spectrum was extracted 
within 3$'$ from the source center, 
while in this analysis, the spectrum was extracted from the 4.5$'$
radius circle including the radio shell.

\subsection{G346.6$-$0.2}

\begin{figure*}
  \begin{center}
    \FigureFile(70mm,70mm){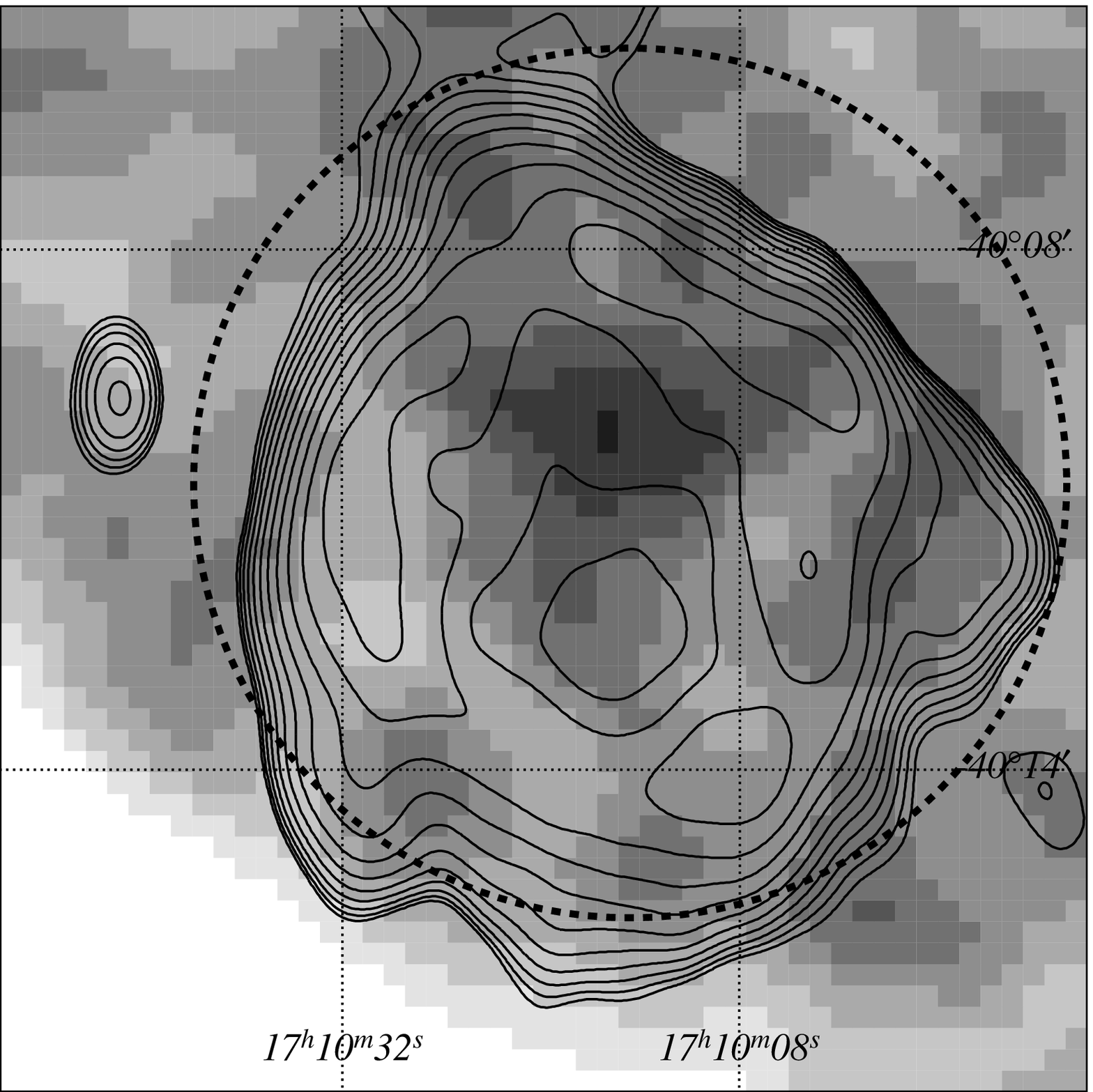}
    \FigureFile(90mm,90mm){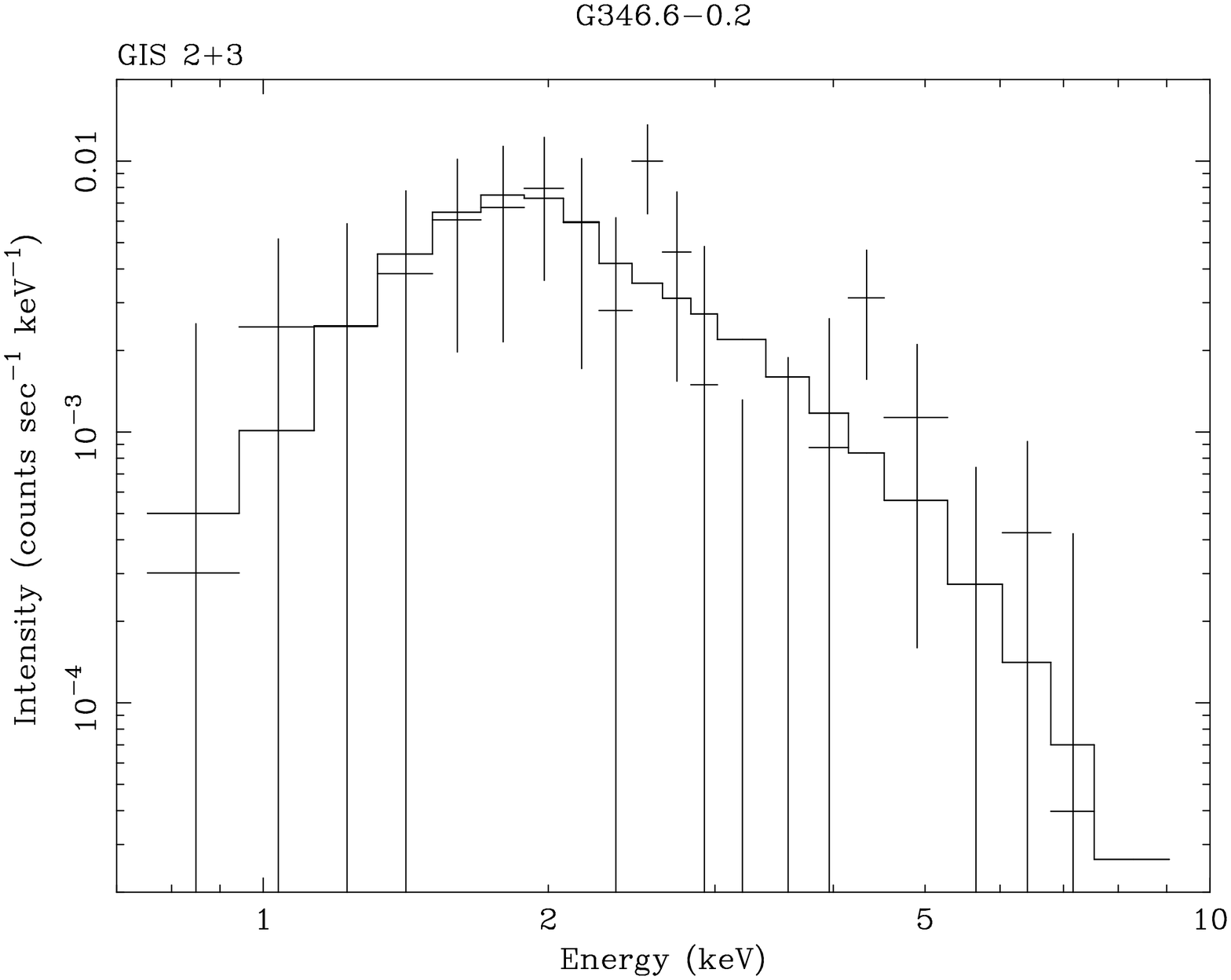}
  \end{center}
  \caption{
Left: GIS 2+3 image of G346.6$-$0.2  obtained in the 0.7--10.0 keV energy band 
(gray scale).
The intensity level are linearly spaced by 0.2 counts pixel$^{-1}$.
The contour shows a radio intensity map at 843 MHz using the 
Molonglo Observatory Synthesis Telescope (MOST) \citep{Whiteoak1996}.
The dotted line shows the region from which the X-ray spectrum was extracted.
Right: GIS spectrum of G346.6$-$0.2 (the crosses) 
and the best-fit power-law model (the histogram).
}\label{fig2}
\end{figure*}

The radio and X-ray images are shown in figure 2.
G346.6$-$0.2 was located near to the edge of the GIS FOV.
The radio shell with a $\sim8'$ diameter was found 
(e.g., \cite{Whiteoak1996}), while
an X-ray image shows an existence of a weak emission in the 
radio shell.

The X-ray spectrum was extracted from a 5$'$ radius circle, while
the background spectrum was extracted from an annular region between 
7$'$ and 5$'$ radius.
The X-ray counts from the source region in the 0.7--10 keV energy band 
were 1258 (GIS2$+$3), 
while the background counts in the same region size and in the same energy 
band were estimated to be 1059 (GIS2$+$3).
The spectrum was explained by either a MEKAL model with a
temperature of 1.6 keV or a single power-law model with a photon
index of 3.7, 
although the errors are large due to the limited photon statistics.
The best-fit power-law model is shown in figure 2.

\subsection{G348.5$+$0.1}
 
\begin{figure*}
  \begin{center}
    \FigureFile(70mm,70mm){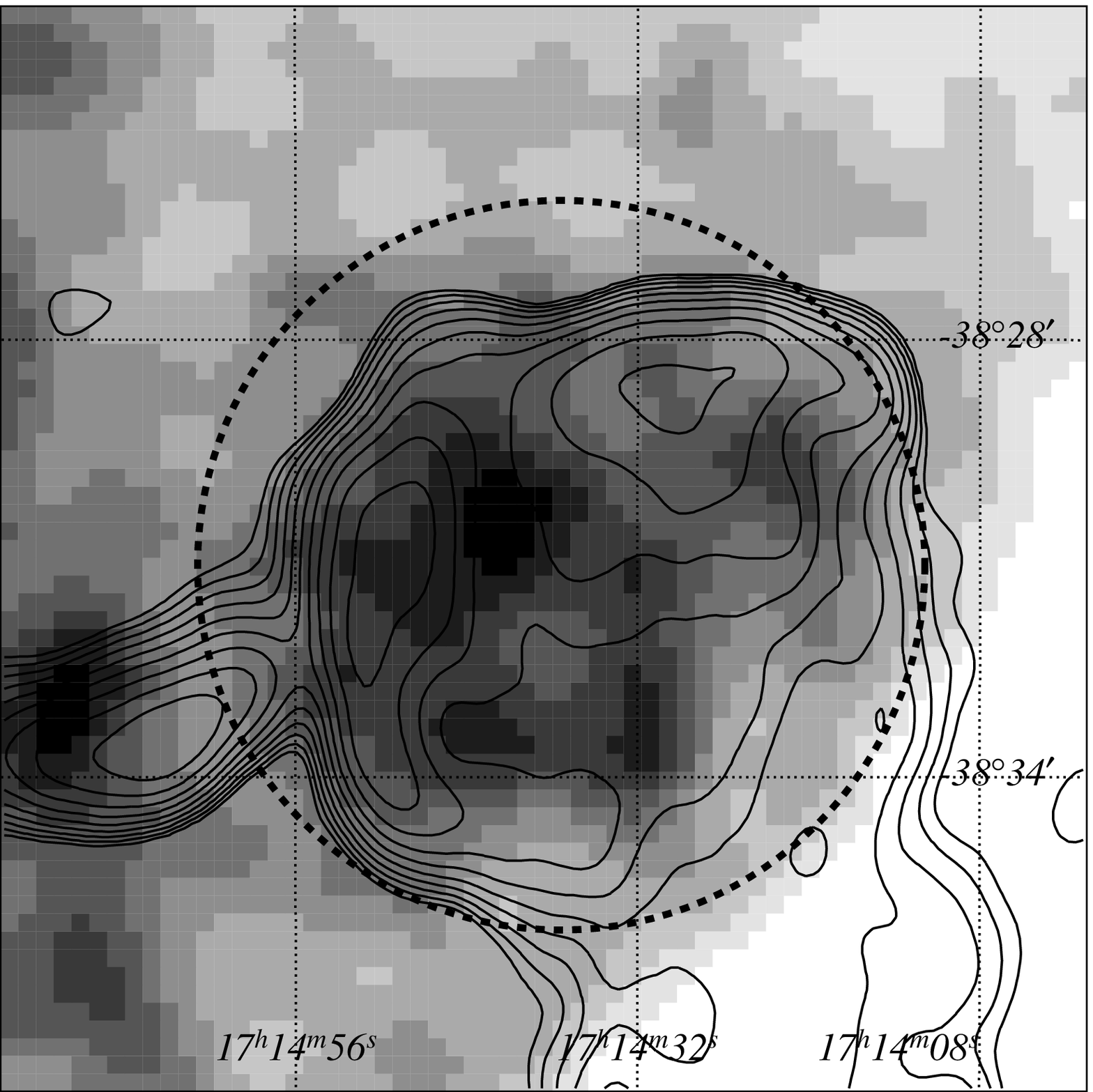}
    \FigureFile(90mm,90mm){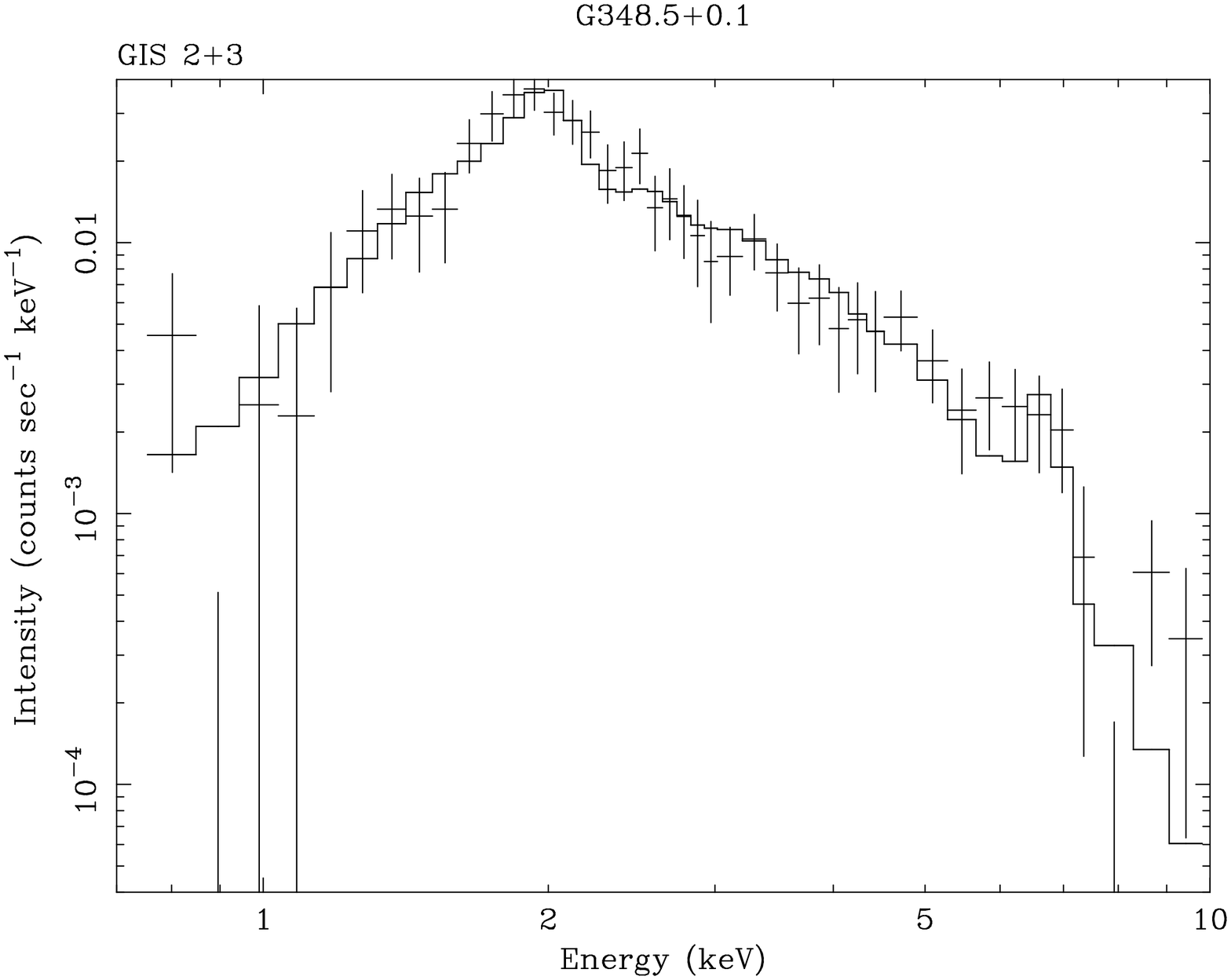}
  \end{center}
  \caption{
Left: GIS 2+3 image of G348.5$+$0.1  obtained in the 0.7--10.0 keV energy band 
(gray scale).
The intensity level are linearly spaced by 0.2 counts pixel$^{-1}$.
The contour shows a radio intensity map by MOST.
The dotted line shows the region from which the X-ray spectrum was extracted.
Right: GIS spectrum of G348.5$+$0.1 (the crosses) and 
the best-fit MEKAL model with freeing the Si abundance (the histogram).
}\label{fig3}
\end{figure*}

The field centered on ($l$, $b$)=(\timeform{348.D7}, \timeform{0.D0}) 
includes 3 radio SNRs, G348.5$-$0.0,
G348.5+0.1, and G348.7+0.3. X-ray emissions from G348.5+0.1 and
G348.7+0.3 were clearly found, but that of G348.5$-$0.0 was 
not clearly seen due to the stray
light from a nearby bright source. 
This data were excluded in the analysis by \citet{Sugizaki2001}.

Figure 3 shows an X-ray image around G348.5$+$0.1. 
G348.5$+$0.1 was located near to the edge of the GIS FOV. 
An extended emission with a diameter of  ${\sim}6'$ 
was clearly found at the position of 
G348.5$+$0.1. The extended X-ray emission well coincides 
with the bright northern 
part of the radio structure (e.g., \cite{Whiteoak1996}). 

The X-ray spectrum was extracted from a 5$'$ radius circle.
Since the stray light from the nearby bright source was found near 
G348.5$+$0.1,
the background spectrum for G348.5$+$0.1 was extracted from the source free
region with a radius of 7.5$'$ 
at the same distance from the Galactic plane and from the FOV center.
The X-ray counts from the source region in the 0.7--10 keV energy band 
were 1483 (GIS2$+$3), 
while the background counts in the same region size and in the same energy 
band were estimated to be 593 (GIS2$+$3).
Due to the poor statistics, the spectrum was well represented by either
a thin thermal emission model with a temperature of 2.2 keV and 
Solar abundances or a power-law
model with a photon index of 3.
Even if we applied a MEKAL model with the Solar abundances,
we found positive residuals at the Si-line band,
which may show that a thin thermal emission is preferable. 
We applied a MEKAL model with freeing the Si abundance to the spectrum.
The best-fit MEKAL model with freeing the Si abundance (model 1Tv in table 3)
is shown in figure 3.

\subsection{G348.7$+$0.3}

\begin{figure*}
  \begin{center}
    \FigureFile(70mm,70mm){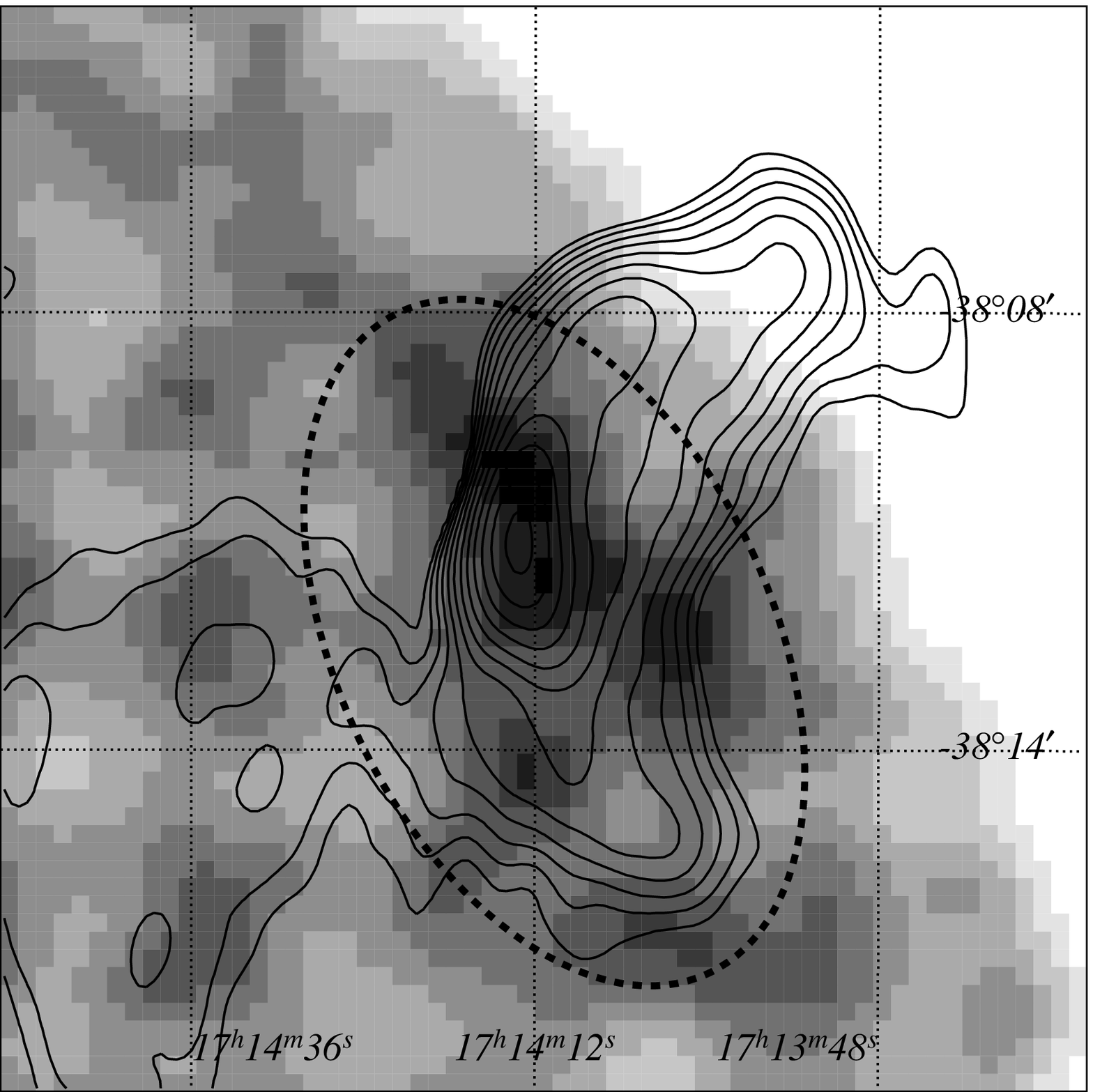}
    \FigureFile(90mm,90mm){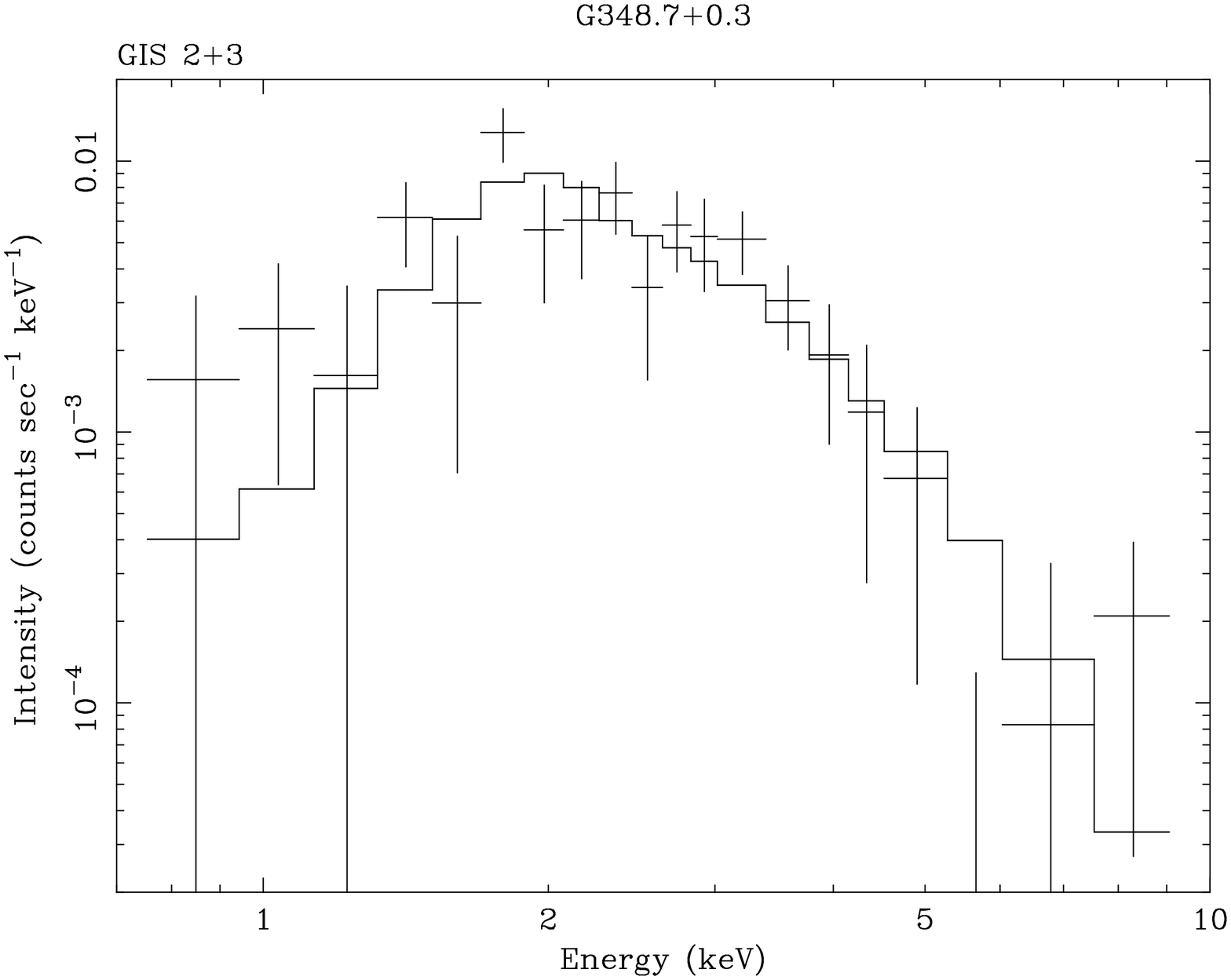}
  \end{center}
  \caption{
Left: GIS 2+3 image of G348.7$+$0.3  obtained in the 0.7--10.0 keV energy band 
(gray scale).
The intensity level are linearly spaced by 0.2 counts pixel$^{-1}$.
The contour shows a radio intensity map by MOST.
The dotted line shows the region from which the X-ray spectrum was extracted.
Right: GIS spectrum of G348.7$+$0.3 (the crosses) and 
the best-fit power-law model (the histogram).
}\label{fig4}
\end{figure*}

We found a source near to the edge of the GIS FOV
(figure 4).
Compared with a radio continuum image (e.g., \cite{Whiteoak1996}),
we found that the X-ray source coincides with a southern part of G348.7$+$0.3.

The X-ray spectrum was extracted from an elliptical region with 
a major axis of 10$'$ and a minor axis of 6$'$.
The background spectrum was extracted from an elliptical region 
surrounding G348.7$+$0.3
with a major axis of 20$'$ and a minor axis of 11$'$, but
excluded the source region.
The X-ray counts from the source region in the 0.7--10 keV energy band 
were 641 (GIS2$+$3), 
while the background counts in the same region size and in the same energy 
band were estimated to be 379 (GIS2$+$3).
Both a MEKAL model ($kT\sim$1.6 keV) and a power-law model ($\alpha\sim4$) 
can explain the X-ray spectrum.
The precise calibration for the outer region 
of the GIS FOV was not established.
Since the SNR was located near to the edge of the GIS FOV, 
the systematic errors of the parameters were large.


\subsection{G355.6$-$0.0}

\begin{figure*}
  \begin{center}
    \FigureFile(70mm,70mm){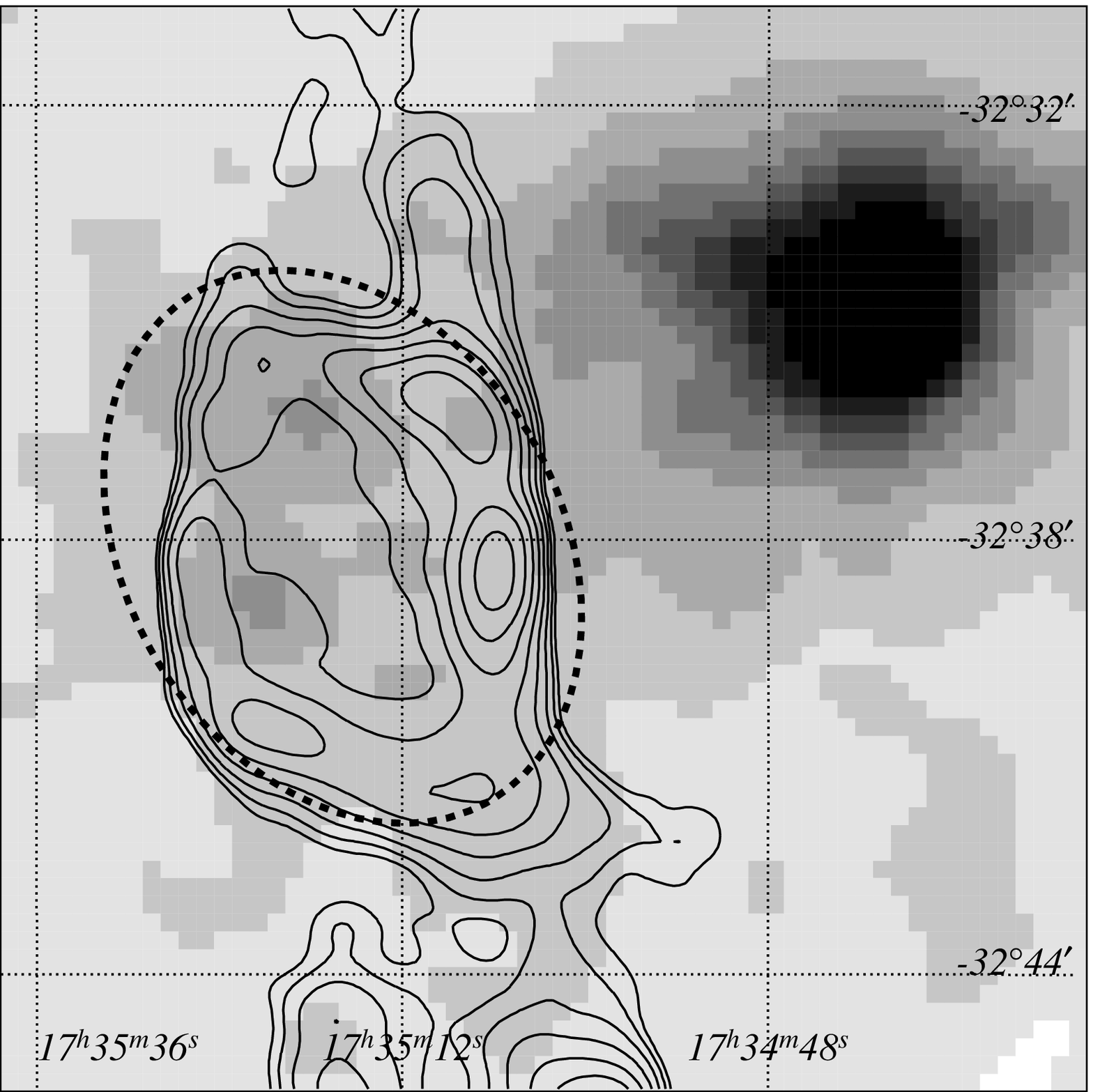}
    \FigureFile(90mm,90mm){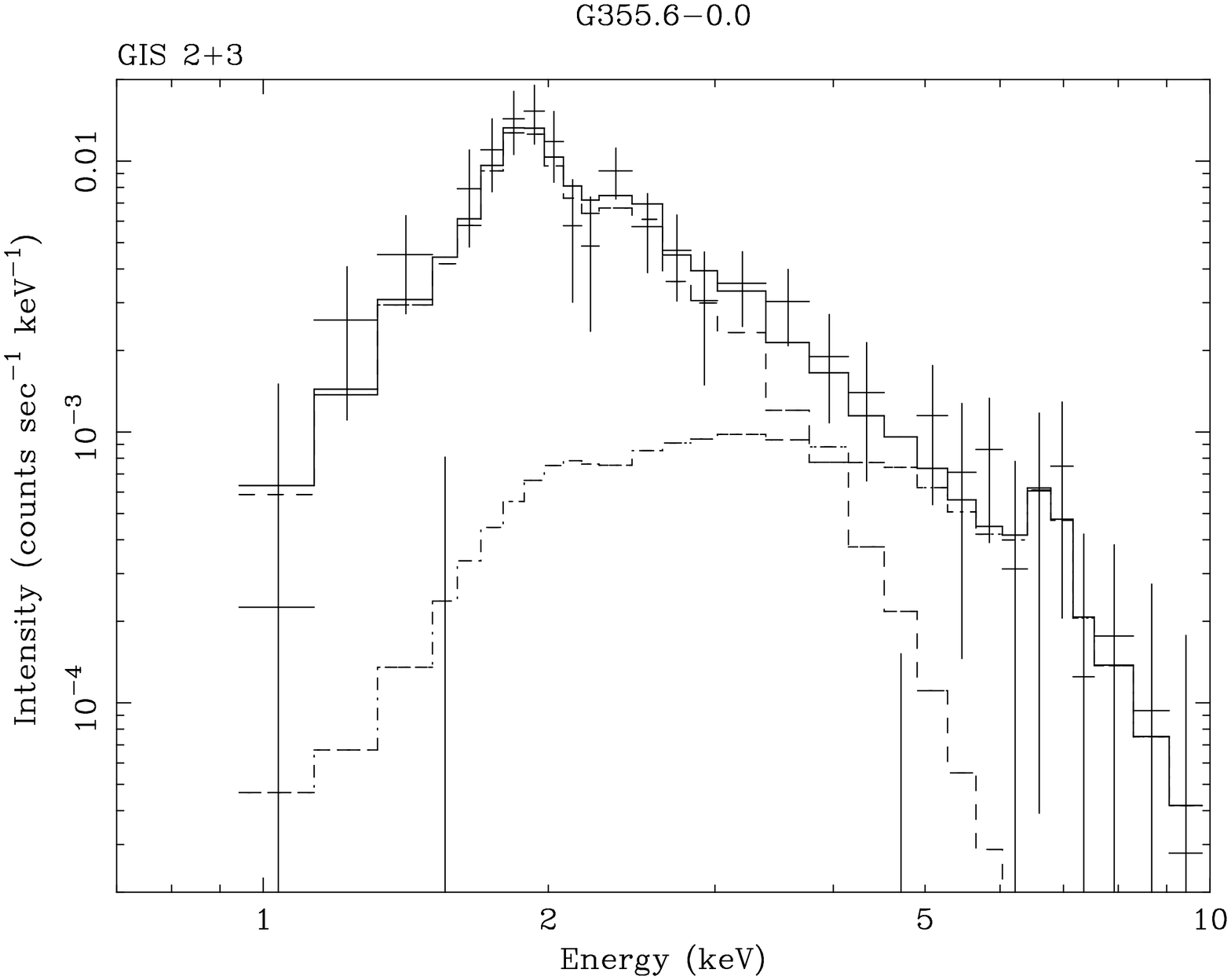}
  \end{center}
  \caption{
Left: GIS 2+3 image of G355.6$-$0.0  obtained in the 0.7--10.0 keV energy band 
(gray scale).
The bright X-ray source in the upper right is a young open cluster NGC 6383.
The intensity level are linearly spaced by 0.4 counts pixel$^{-1}$.
The contour shows a radio intensity map by NVSS.
The dotted line shows the region from which the X-ray spectrum was extracted.
Right: GIS spectrum of G355.6$-$0.0 (the crosses) and 
the best-fit 2 MEKAL model (the histogram).
The each component is shown by the dashed line and the dash-dotted line.
}\label{fig5}
\end{figure*}

X-ray detection of G355.6$-$0.0 
in AGPS 
and results of a power-law model fit were 
reported by \citet{Sugizaki2001}.
A bright X-ray source in the image (figure 5) is a young open cluster NGC 6383.
G355.6$-$0.0 was located at the east of NGC 6383. 
An extended X-ray emission  
was found at the position of G355.6$-$0.0.
The size was comparable to that of the radio band 
($\sim8'\times6'$, e.g., \cite{Condon1998}) 

The X-ray spectrum was extracted from an elliptical region with 
a major axis of 8$'$ and a minor axis of 6$'$.
The background spectrum was extracted from an annular region between 
10$'$ and 5$'$ radius, but the contribution from NGC 6383
were excluded.
The X-ray counts from the source region in the 0.7--10 keV energy band 
were 775 (GIS2$+$3), 
while the background counts in the same region size and in the same energy 
band were estimated to be 423 (GIS2$+$3).
The results of the spectral fit are listed in table 3.
Taking account of the errors,
we found that the power-law model fit are consistent with those in 
\citet{Sugizaki2001}.
The X-ray spectrum of G355.6$-$0.0 exhibited an emission line from
Si (figure 5), which suggested that it is thermal plasma emission.
Although the photon statistics are limited, the X-ray spectrum inferred the 
presence of a Fe K-line emission.
When a single-temperature MEKAL model was applied,
systematic residuals were found above 5 keV energy band.
If we let metal abundance be free,
high Si and Fe abundances were required and positive residuals in the high
energy band remained.
This indicates the presence of the additional high-temperature 
component.
Thus, we applied a two-temperature MEKAL model and obtained an acceptable fit
with temperatures of 0.7$^{+0.3}_{-0.4}$ keV and 7.5 ($>$1.7) keV 
and Solar abundances.

\section{Discussion and Conclusions}

AGPS covered all the Galactic 
inner disk ($| l | < 45^{\circ}$ and $| b | < 0.4^{\circ}$) and the 
Galactic center region ($| l |<2^{\circ}$ and $| b |<2^{\circ}$) and
detected X-ray emissions from many X-ray sources 
embedded in the Galactic plane, 
which provided us with new information on their X-ray emission.
AGPS detected X-rays from $\sim30$ cataloged SNRs, including
15 SNRs undetected in the X-ray band before AGPS.
We presented X-ray images and X-ray spectra of 5 SNRs among them, 
G12.0$-$0.1, G346.6$-$0.2, G348.5$+$0.1,G348.7$+$0.3, and G355.6$-$0.0.

The sizes of X-ray emissions from G12.0$-$0.1, G348.5$+$0.1, G348.7$+$0.3,
and G355.6$-$0.0 are comparable to their radio structures,
while that of G346.6$-$0.2 is smaller than the radio structure.
All of the X-ray spectra were heavily absorbed by interstellar matter 
($N_{\rm H}>10^{22}$ cm$^{-2}$) and their X-ray fluxes 
in the 0.5--10.0 keV energy band were in the 
range of $\sim$10$^{-12}$--10$^{-11}$ erg s$^{-1}$ cm$^{-2}$.
The Galactic plane contains a large amount of absorbing matters 
and many bright X-ray binary sources.
The large $N_{\rm H}$ value 
and low flux of the SNRs (probably due to long distances to the SNRs)
are the reason why their X-ray emissions could not be
detected before AGPS.

Recently, Aharonian et al. (2008a, 2008b) reported 
detection of TeV ${\gamma}$-ray emission from G348.5$+$0.1 and G348.7$+$0.3 
with High Energy Stereoscopic System (H.E.S.S.)
(HESS J1714$-$385 and HESS J1713$-$381, respectively) 
and results of follow up X-ray observations of the SNRs 
with Chandra and XMM-Newton.  
They resolved the X-ray emission detected in AGPS into
an extended thermal emission and several point-like sources and discussed  
the nature of the X-ray and $\gamma$-ray sources in detail.
The X-ray spectra of G348.5$+$0.1 and G348.7$+$0.3 obtained in AGPS 
contain X-ray emissions from the extended source and the point-like sources. 
We confirmed that the ASCA spectra of G348.5$+$0.1 and G348.7$+$0.3 were 
well represented by a model consisting of X-ray emissions from 
the extended and point-like sources 
reported in Aharonian et al. (2008a, 2008b).

Here, we briefly discuss on the X-ray emission 
from the other three SNRs, G12.0$-$0.1, G346.6$-$0.2, and G355.6$-$0.0.
The spectrum of G355.6$-$0.0 exhibited a Si-K line 
(and possibly a Fe-K line), 
which indicates that the X-ray emission has a thin thermal plasma origin.
On the other hand, 
no clear emission line features were found in the spectra of the others
and their spectra could be represented by either a thin thermal emission model 
or a power-law model.


Although errors are large, the temperatures of G12.0$-$0.1, 
G346.6$-$0.2, and G355.6$-$0.0 derived from the MEKAL model 
fits (1--2 keV) are consistent with those of SNRs.
Assuming that the SNRs are in the Sedov phase \citep{Sedov1959},
we examined a case of a shell-like SNR with thin thermal X-ray emission.
The size of the SNR is derived from the radio and X-ray images, while
the plasma temperature ($kT_{\rm X}$) and the emission integral 
($EI$=$\int n_{\rm H}n_{\rm e}dV$, $n_{\rm H}$ and $n_{\rm e}$
are the hydrogen and electron densities, respectively, $V$ is the
volume of the X-ray emitting region) are taken from the MEKAL model fit.
The distances to the SNRs were calculated from the best-fit $N_{\rm H}$ 
values with assumption of the mean density of 1 Hcm$^{-3}$.
For G355.6$-$0.0, we assumed the low-temperature component 
is X-ray emission from the shell region.
Assuming the temperature of the SNR shell ($kT_{\rm S}$) of 
$kT_{\rm S}=0.775\ kT_{\rm X}$, 
the filling factor of 0.25, and $n_{\rm e}$=1.2\ $n_{\rm H}$,
the explosion energies for G12.0$-$0.1, G346.6$-$0.2, 
and G355.6$-$0.0 were calculated to be $\sim$1, $\sim$0.05, and $\sim$0.5 
in units of 10$^{51}$ erg, respectively.
Although the uncertainties of the physical parameters derived 
from the spectral fits are large, 
the case of G346.6$-$0.2 seems to be much smaller than the 
canonical supernova explosion energy (=10$^{51}$ erg).


Next, we examined two cases of nonthermal X-ray emission from the SNR 
for G12.0$-$0.1 and G346.6$-$0.2, 
a SN 1006-like case and a pulsar wind nebula (PWN) case.
In the case of SN 1006-like SNRs, 
high energy electrons accelerated in the SNR shells emit 
synchrotron X-rays and lose their energy.
The observed flux densities at 10$^{18}$ Hz were 
$\sim$9$\times$10$^{-8}$ Jy and $\sim$7$\times$10$^{-8}$ Jy
for G12.0$-$0.1 and G346.6$-$0.2, respectively, 
which are smaller than the values extrapolated from the radio band
with the parameters in table 1 \citep{Green2006}, 
$\sim$(1--2)$\times$10$^{-6}$ Jy and $\sim$3$\times$10$^{-4}$ Jy 
for G12.0$-$0.1 and G346.6$-$0.2, respectively.
Furthermore, 
although the errors are large, 
the spectral indices in the X-ray band obtained 
from the power-law model fit were steeper than those in the radio band. 
These suggest that 
there is a break frequency between the radio and the X-ray bands.
 Therefore, the nonthermal X-rays from the SNR shells like SN 1006,
may be possible.
The X-ray morphology of G346.6$-$0.2 seems 
to show a weak peak in the radio structure.
Although a pulsar has not been discovered  
for G346.6$-$0.2, a PWN scenario can not be excluded.

Unfortunately, detailed properties of the SNRs are still unknown
because the photon statistics in the AGPS data were limited.
In order to understand the physical process in the SNRs and their nature,
spatially resolved X-ray spectra with good photon statistics 
with Chandra, XMM-Newton, and Suzaku are required.

\vspace{1pc}

We would like to express our thanks to all of the ASCA team. 
This work is supported in part by the Grant-in-Aid for Scientific
Research 
from the Japan Society for the Promotion of Science (JSPS)
(Nos. 15540225 and 18540228; S.Y.).
M. U. and A. B. are supported by Research Fellowships for 
Young Scientists of JSPS.


\end{document}